\pdfoutput=1
\documentclass[prl,twocolumn,floatfix]{revtex4}

\usepackage{amsmath, amssymb}
\usepackage{graphicx}		
\usepackage{bm}					
\usepackage{array}

\def\avg#1{\ensuremath{\left< #1 \right>}}

\newcommand{\bra}[1]{\ensuremath{\left\langle #1\right|}}
\newcommand{\ket}[1]{\ensuremath{\left|#1\right\rangle}}

\newcommand{\highstate}{1}
\newcommand{\lowstate}{0}
\newcommand{\up}{\ket{\highstate}}
\newcommand{\down}{\ket{\lowstate}}
\newcommand{\uu}{\ket{\highstate\highstate}}
\newcommand{\ud}{\ket{\highstate\lowstate}}
\newcommand{\du}{\ket{\lowstate\highstate}}
\newcommand{\dd}{\ket{\lowstate\lowstate}}

\newcommand{\Be}{\ensuremath{{^9}{\rm Be}^{+} \,}}
\newcommand{\Mg}{\ensuremath{{^{24}}{\rm Mg}^{+} \,}}

\newcommand{\tr}{\ensuremath{{\textrm T}}}
\newcommand{\figcircuit}{1}
\newcommand{\figstates}{2}
\newcommand{\figfidelities}{3}

\begin{document}

\title{Realisation of a programmable two-qubit quantum processor}


\author{D. Hanneke, J. P. Home, J. D. Jost, J. M. Amini, D. Leibfried \& D. J. Wineland}

\affiliation{Time and Frequency Division, National Institute of Standards and Technology, Boulder, Colorado 80305, USA}

\begin{abstract}
The universal quantum computer\cite{deutschProcRoyalA1985} is a device capable of simulating any physical system\cite{feynmanIJTP1982} and represents a major goal for the field of quantum information science. Algorithms performed on such a device are predicted to offer significant gains for some important computational tasks\cite{NielsenChuang2000}
. In the context of quantum information, ``universal'' refers to the ability to perform arbitrary unitary transformations in the system's computational space\cite{deutschProcRoyalA1989}. The combination of arbitrary single-quantum-bit (qubit) gates with an entangling two-qubit gate is a gate set capable of achieving universal control of any number of qubits\cite{barencoPRA1995,bremnerPRL2002,zhangPRL2003}, provided that these gates can be performed repeatedly and between arbitrary pairs of qubits. Although gate sets have been demonstrated in several technologies\cite{southwellNature2008}, they have as yet been tailored toward specific tasks, forming a small subset of all unitary operators. Here we demonstrate a programmable quantum processor that realises arbitrary unitary transformations on two qubits, which are stored in trapped atomic ions. Using quantum state and process tomography\cite{hradilSpringer2004}, we characterise the fidelity of our implementation for 160 randomly chosen operations. This universal control is equivalent to simulating any pairwise interaction between spin-1/2 systems. A programmable multi-qubit register could form a core component of a large-scale quantum processor, and the methods used here are suitable for such a device\cite{homeScience2009}.
\end{abstract}

\maketitle

Computers are useful because they are versatile. Changing the problem to be solved amounts to reconfiguring inputs to the processor, that is, to reprogramming it. In a classical computer, a program is ultimately decomposed into sequences of operations implemented with logic gates. The explosion of interest in quantum information science coincided with the realisation that a similar decomposition exists for quantum processors\cite{deutschProcRoyalA1989,barencoPRA1995,bremnerPRL2002}; arbitrary operations on a multi-qubit system can be broken down into sequences of discrete operators -- ``quantum gates''. As with its classical counterpart, a programmable quantum computer is more versatile than one designed for a fixed task.

Ease of implementation can favor certain decompositions of quantum operations, for example, those based on arbitrary single-qubit gates and a single entangling two-qubit gate\cite{barencoPRA1995,bremnerPRL2002}. Since realising gates acting on two or more qubits tends to be more experimentally challenging\cite{southwellNature2008}, much attention has been focused on using them optimally in the creation of entanglement\cite{krausPRA2001} and on finding decompositions minimizing the number of times they are applied\cite{zhangPRA2003,zhangPRL2003,vidalPRA2004,vatanPRA2004,shendePRA2004}. Some well-chosen operations can be performed with two or fewer applications of two-qubit gates, but these form an infinitesimal subset in the space of two-qubit operations\cite{vidalPRA2004}. For a two-qubit system with a maximally entangling two-qubit gate, three applications of this gate, when combined with additional applications of arbitrary single-qubit gates, are sufficient for universality\cite{zhangPRA2003}. Here we present and characterise a universal quantum processor that can produce any desired two-qubit unitary transformation when programmed with 15 classical inputs\cite{krausPRA2001,vidalPRA2004}.

The decomposition of a given operation depends on the available gate sets. Our choice of a universal gate library consists of single-qubit gates and one maximally entangling two-qubit gate. The single qubit gates are rotations
\begin{align}
	R\!\left( \theta,\phi \right) &= \exp\!\left[-i\theta\left(\cos\phi~\sigma_x + \sin\phi~\sigma_y\right)/2\right]
																			\\
	R_z\!\left(\phi_z\right) 						 &= \exp\!\left(-i\phi_z\sigma_z/2\right) 
\end{align}		
in the computational basis $\up \equiv (1,0)$, $\down \equiv (0,1)$. Here, $\sigma_x, \sigma_y, \sigma_z$ are the Pauli matrices. The single-qubit gates have variables $\theta, \phi, \phi_z$ that can take any value from 0 to $2\pi$. The entangling two-qubit gate is
\begin{equation}							
	G = e^{-i\pi/4} \exp\!\left(\frac{i \pi}{4}\sigma_z\otimes\sigma_z\right)
\end{equation}
and operates on the basis $\uu,\ud,\du,\dd$. Here, $\otimes$ indicates the tensor product, and $\ket{ij} \equiv \ket{i}\otimes\ket{j}$.

\begin{figure}
	\begin{center}
	\includegraphics[width=\columnwidth]{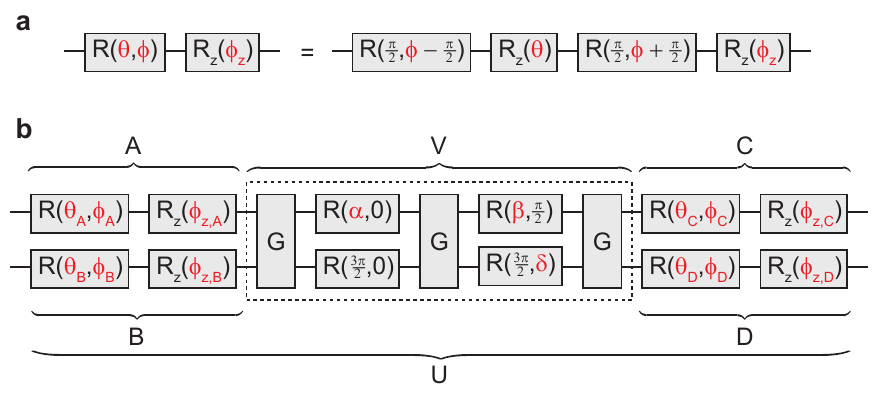}
	\end{center}
	\caption{\textbf{Components for universal computation.} Circuit diagrams for arbitrary unitary transformations on \textbf{a,} one and \textbf{b,} two qubits. The operations for each circuit are implemented from left to right with each line representing one qubit. Part \textbf{a} also indicates the decomposition employed for $R(\theta,\phi)$. The dashed box in part \textbf{b} contains the three degrees of freedom $\alpha, \beta, \delta$ that determine the two-qubit operation's local equivalence class. The brackets highlight the decomposition of the two-qubit operation $U$ as described in the text: $U = (C \otimes D) \cdot V \cdot (A \otimes B)$.}
\end{figure}

With this gate library, the circuits shown in Figure~\figcircuit~can be used to implement arbitrary unitary transformations on (a) one qubit and (b) two qubits. The single-qubit operation is characterized by three degrees of freedom and may be decomposed into the matrix product $R_z(\phi_z) \cdot R(\theta,\phi)$. Each two-qubit unitary transformation is described by 15 degrees of freedom\cite{krausPRA2001,vidalPRA2004}. The group of all such transformations can be divided into subsets that are equivalent up to single-qubit operations. Such subsets are called local equivalence classes\cite{makhlinQIP2002} because local operations can map among all members of the class. Each local equivalence class can be described by three parameters\cite{krausPRA2001}. Given a unitary transformation $U$, we decompose it into $U = (C \otimes D) \cdot V \cdot (A \otimes B)$. Here, $V$ is in the same local equivalence class as $U$ and is in a special form that requires fewer gates on our processor. $A$ and $C$ are single-qubit operations on one of the qubits, and $B$ and $D$ are single-qubit operations on the other.

Determination of the 15 single-qubit-gate parameters is facilitated by working in the so-called ``magic'' basis\cite{krausPRA2001}:
\begin{equation}
\begin{split}
	\frac{1}{\sqrt{2}}\left(\uu+\dd\right), \frac{i}{\sqrt{2}}\left(\uu - \dd\right),\\
	\frac{i}{\sqrt{2}}\left(\ud+\du\right), \frac{1}{\sqrt{2}}\left(\ud-\du\right).
\end{split}
\end{equation}
This basis amounts to the Bell states with specific global phases, and we take advantage of two of its convenient mathematical properties. First, single-qubit operations with unit determinant are given in this basis by real matrices that are orthogonal\cite{makhlinQIP2002}. Second, two-qubit operations $u$ and $v$ in SU(4) are in the same local equivalence class if and only if $u u^\tr$ and $v v^\tr$ have the same eigenvalues\cite{makhlinQIP2002} ($u^\tr$ is the matrix transpose of $u$, and we use lower-case letters to represent matrices in the magic basis). The decomposition of a given operation $U$ follows a two-step procedure analogous to that in Ref.~\cite{shendePRA2004}. (See Methods for details and Supplementary Information for examples.) Briefly, we first match matrix eigenvalues to find a special element $V$ in the same local equivalence class as $U$ (i.e. find $\alpha$, $\beta$, $\delta$, the three degrees of freedom in Figure~\figcircuit b's dashed box). Second, we manipulate real, orthonormal matrix eigenvectors to find the four remaining single-qubit operations required to map between $V$ and $U$. Because the magic basis properties rely on unit matrix determinants, we can implement operations only up to a global phase. Global phases exactly vanish in any observable quantity, so this restriction has no physical relevance.

We implement the quantum circuit with trapped ions using techniques applicable for scaling to a larger system\cite{homeScience2009}. Each qubit is stored in a pair of energy eigenstates in the $\Be 2s\,^2S_{1/2}$ hyperfine manifold. The qubit basis states can be transferred between different pairs of the eight hyperfine levels\cite{homeScience2009}. The qubit spends the majority of its time stored in the $(\up,\down) = (\ket{F=1,m_F = 0}$, $\ket{F=2, m_F=1})$ ``magnetic-field-independent'' manifold, for which the energy splitting has zero first-order dependence on the magnetic field at our chosen value of $0.011964$~T, leading to long coherence times (15~s has been measured for a pair of states with similar second-order field dependence\cite{langerPRL2005}).

The two \Be ions are stored simultaneously with two \Mg ions in a six-zone linear Paul trap\cite{barrettNature2004,jostNature2009,homeScience2009}; the ions form a linear chain along the axis of weakest confinement. Coulomb repulsion couples the motion of all four ions such that laser cooling the \Mg ions sympathetically cools all of the ions without destroying the quantum information\cite{jostNature2009,homeScience2009} stored in the \Be ions. The collective motion of the four ions can be described as the sum of 12 normal modes, four along each of the principal axes. 
The two-qubit gate uses two modes involving motion along the axis of the ion chain\cite{jostNature2009,homeScience2009}. We spectrally address the \Mg ions to Doppler and resolved-sideband cool these modes to near the quantum ground state of motion\cite{homeScience2009}. Since the ion spatial order affects the mode frequencies, and because both resolved-sideband cooling and the two-qubit gate require spectral addressing of the modes, we deterministically initialize the ion order\cite{jostNature2009} to \Be--\Mg--\Mg--\Be at the beginning of each experimental sequence.

State preparation and measurement are performed using resonant laser light that couples the \Be $2s\,^2S_{1/2}$ states to the $2p\,^2P_{1/2}$ and $2p\,^2P_{3/2}$ states\cite{winelandJRNIST1998}. Projective measurements in the single-qubit computational basis $\up,\down$ utilize a cycling transition\cite{winelandJRNIST1998,homeScience2009}. Measurements in other bases are made by first rotating their eigenvectors into the computational basis. A single detection apparatus sequentially measures the two qubits independently\cite{homeScience2009}.

The universal gate set above is implemented with laser-induced stimulated-Raman transitions\cite{winelandJRNIST1998}. The two-qubit gate $G$ is a geometric phase gate\cite{leibfriedNature2003,homeScience2009}. The single-qubit gate $R(\theta,\phi)$ can be produced by driving resonant Rabi oscillations between the qubit states, where the angle $\theta$ is controlled by the laser pulse intensity and duration. The phase $\phi$ is set by the phase difference between the two Raman light fields at the ion relative to the qubit phase\cite{winelandJRNIST1998}. It is controlled via the phase of a radio-frequency (RF) potential applied to an acousto-optic modulator (AOM). The single-qubit gate $R_z(\phi_z)$ advances the qubit phase by $\phi_z$ relative to that of the Raman light fields. It is implemented by retarding the phase of an AOM's RF for subsequent laser pulses. In order to individually apply $R(\theta,\phi)$ to each qubit, time-dependent electric potentials\cite{roweQIC2002} divide the four-ion chain into two \Be--\Mg pairs and transport them to zones separated by 240~$\mu$m. The four applications of $R(\theta,\phi)$ per qubit in Figure~\figcircuit b require four such ion separations and subsequent recombinations -- a total information transport of nearly 2~mm per qubit. An AOM placed before the trap directs the laser beams to the relevant ion.

\begin{figure*}
	\begin{center}
	\includegraphics[width=\textwidth]{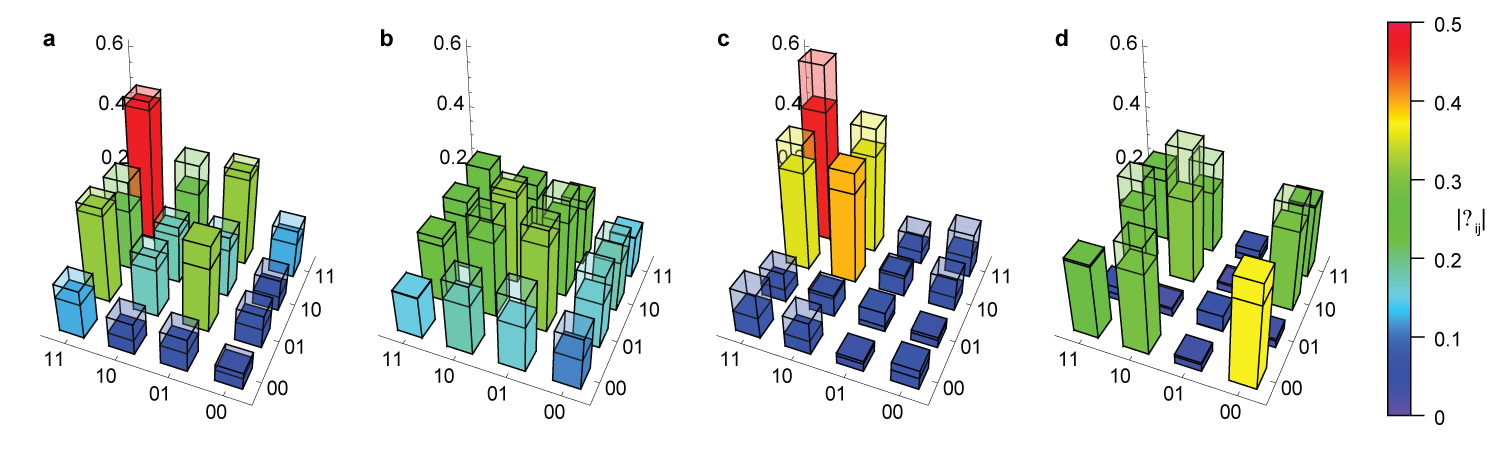}
	\end{center}
	\caption{\textbf{Diversity of states.} Examples of the density matrices created by applying four of the randomly chosen unitary transformations to simple product states. The bar heights and colors show the absolute value of the output density matrices, with the experimental data shown solid and the theory shown semi-transparent. These states have fidelities $f$ = 0.81, 0.80, 0.75, 0.80, relative to the ideal states. The Supplementary Information gives the input states and operations used in these examples.}
\end{figure*}

The requirement that single-qubit-gate inputs take any value from 0 to $2\pi$ prompts a further decomposition of $R(\theta,\phi)$. The RF potentials that control $\phi$ and $\phi_z$ are generated by a stable oscillator whose phase is easily controllable. The variable $\theta$, however, depends on the laser pulse's intensity, which for technical reasons is not constant for the duration of the pulse. Rather than calibrate this for arbitrary $\theta$, we calibrate a single value, $\theta=\pi/2$, and decompose $R(\theta,\phi)$ into
\begin{equation}
	R\!\left( \theta,\phi \right) = R\!\left(\frac{\pi}{2},\phi+\frac{\pi}{2}\right)
									\cdot R_z\!\left(\theta\right) \cdot R\!\left(\frac{\pi}{2},\phi-\frac{\pi}{2}\right).
\end{equation}
In this way, all 15 inputs to our universal circuit are controlled by shifting the phase of a control oscillator relative to the qubit. The number, duration, and spacing of the laser pulses are identical for every $U$.

To demonstrate the ability of the processor to generate arbitrary unitary transformations, we program it with 160 different randomly chosen operations distributed in SU(4) according to the Haar measure\cite{mezzadriNoticesAMS2007}. (The probability distribution given by the Haar measure is a uniform distribution in the space of unitary matrices.) To characterise our implementation of the 160 operations, we apply each to one of 16 input states formed by the tensor products of $\down$, $\up$, $\ket{+}\equiv\left(\up + \down\right)/\sqrt{2}$, and $\ket{-i}\equiv\left(\up - i \down\right)/\sqrt{2}$. The assignment of operation to input state is random with the constraint that all input states are used an equal number of times. The application of an operation to its input state produces an output density matrix, which we reconstruct using quantum state tomography\cite{roosPRL2004,hradilSpringer2004}. This procedure involves nine analysis settings. For each setting, we run the experimental sequence 100 times for a total of 900 runs per unitary transformation. A single run takes approximately 37~ms. We compare the measured output state to the ideal result using the fidelity\cite{jozsaJModOptics1994}, 
$f(\rho_\textrm{ideal},\rho_\textrm{exp}) \equiv \left[\textrm{Tr}\left(\sqrt{\sqrt{\rho_\textrm{ideal}}\rho_\textrm{exp}\sqrt{\rho_\textrm{ideal}}} \,\right)\right]^2$. Figure~\figstates ~shows four examples of the output states, and Figure~\figfidelities a gives a histogram of the 160 state fidelities. The 160 operations have a mean state fidelity of $\avg{f} = 79.1(4.5) \,\%$, where the error bar is the standard deviation of the measurements. Numerical estimates\cite{homeScience2009} indicate that 3.4\,\% of this distribution arises from statistical fluctuations in photon counts used in state measurement. We attribute the remaining distribution to variability in each operation's susceptibility to experimental noise. We observe no correlation between output state fidelity and input state, as demonstrated in Figure~\figfidelities b,c. The mean output state fidelities from operating on the 16 input states are distributed with a standard deviation of 1.5\,\%, as we would expect for the means of 10 measurements which themselves have a standard deviation of 4.5\,\%. The primary fidelity loss mechanisms are percent-level intensity fluctuations in the Raman light fields\cite{homeScience2009} and spontaneous emission\cite{ozeriPRA2007}; the fidelities observed here are consistent with those demonstrated previously\cite{homeScience2009} after accounting for the increased number of gates.

\begin{figure}
	\begin{center}
	\includegraphics[width=\columnwidth]{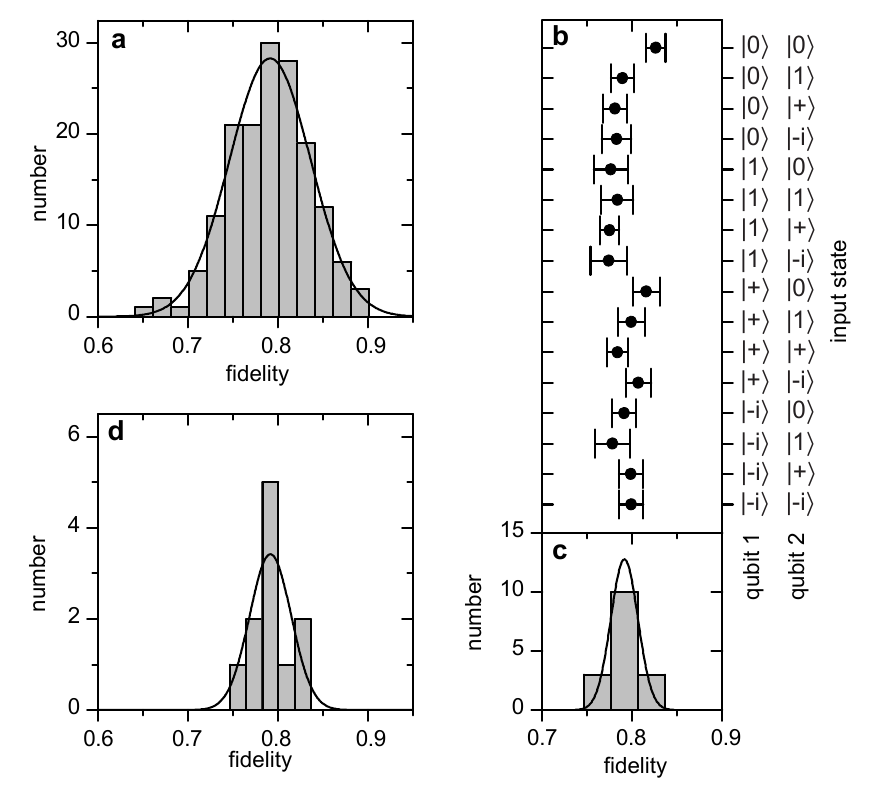}
	\end{center}
	\caption{\textbf{Characterization of arbitrary control.} \textbf{a,} Histogram of the output state fidelities $f$ for 160 arbitrary unitary operations applied to one of 16 input product states. The mean output state fidelities $\avg{f}$ for each input state \textbf{b,} plotted showing the standard error of the mean and \textbf{c,} binned into a histogram. \textbf{d,} The mean state fidelities $\bar{f}$ for the 11 operations reconstructed using quantum process tomography. The curves are normal distributions with the same respective mean and standard deviation as the data. The narrower distributions in histograms \textbf{c,d} are expected because they are distributions of means.}
\end{figure}

As a further check on input-state independence, we conduct quantum process tomography\cite{hradilSpringer2004,homeScience2009} on 11 of the operations. Process tomography reconstructs the completely positive linear map $\mathcal{E}$ that describes the qubit evolution from initial to final density matrices,
$\rho_\textrm{final} = \mathcal{E}(\rho_\textrm{initial})$.
The map includes the possibility of experimental imperfections such as coupling to the environment, which leads to nonunitary evolution. We represent the map by a $16\times16$ matrix\cite{hradilSpringer2004}
$E=\sum_{i,j}\ket{i}\bra{j}\otimes\mathcal{E}(\ket{i}\bra{j})$.
For each of the 11 operations, we determine an experimental process matrix $E_\textrm{exp}$ and compare it with the ideal case by calculating both the entanglement fidelity\cite{horodeckiPRA1999}
$F \equiv \textrm{Tr}\left(E_\textrm{ideal} E_\textrm{exp}\right)/16$
and the mean state fidelity $\bar{f}$ obtained by averaging the output state fidelities from numerically applying $E_\textrm{exp}$ and $E_\textrm{ideal}$ to an unbiased set of 36 input states (formed from the eigenstates of tensor products of the Pauli matrices). These fidelities are related by $\bar{f} = (F d+1)/(d+1)$, where $d$ is the Hilbert-space dimension\cite{horodeckiPRA1999} (here $d=4$). The 11 processes have mean fidelities of $\avg{F} = 73(3) \,\%$ and $\avg{\bar{f}} = 79(2) \,\%$.

In conclusion, we have demonstrated a programmable quantum processor capable of implementing all possible unitary operations on two qubits. To address large-scale problems, many more qubits and gates will be required. In anticipation of such applications, this implementation used only scalable techniques\cite{homeScience2009} such as long-lived qubit storage, quantum information transport, and sympathetic cooling. When implementing a larger system, the compound errors from successive operations will need to be reduced via error correction\cite{NielsenChuang2000}. This will require much higher gate fidelities than shown here, both to achieve fault-tolerance and to reduce error correction's computational overhead\cite{knillNature2005}. Nevertheless, the type of device described here could form a processing unit in a larger system\cite{southwellNature2008} with programmable registers connected by multidimensional trap arrays\cite{winelandJRNIST1998,kielpinskiNature2002} or photonic networks\cite{duanAAMOP2008}.

\section*{Methods}

\subsection{Algorithmic details.}

For a given two-qubit operation $U$, calculating the 15 single-qubit-gate parameters used in the circuit of Figure~\figcircuit b is facilitated by working in the so-called ``magic'' basis\cite{hillPRL1997,krausPRA2001} given in the main text. Transforming to the magic basis from the two-qubit computational basis $\uu, \ud, \du, \dd$ is accomplished by use of the unitary matrix
\begin{equation}
	\Lambda = \frac{1}{\sqrt{2}}\begin{pmatrix} 1 & i & 0 & 0 \\
																							0 & 0 & i & 1 \\
																							0 & 0 & i & -1 \\
																							1 &-i & 0 & 0 \end{pmatrix}.
\end{equation}
The properties of the magic basis rely on unit matrix determinants; thus we first strip $U$ of any global phase by dividing it by a fourth-root of its determinant, making it a member of SU(4). Global phases exactly vanish in any observable quantity, allowing this modification. In what follows, matrices in the computational basis are denoted with capital letters, and those in the magic basis by lower-case letters; e.g., $m=\Lambda^\dagger M \Lambda$.

We first find the three degrees of freedom $\alpha$, $\beta$, $\delta$ that produce the correct local equivalence class. We decompose the circuit in Figure~\figcircuit b as 
\begin{equation}
	U=\left(C \otimes D\right) \cdot V \cdot \left(A \otimes B\right),
	\label{eq:computationalcircuit}
\end{equation}
where $V$ determines the local equivalence class and can be generated using the gate operations appearing in Figure~\figcircuit b within the dashed box. In order to construct $V$, we transform both $V$ and $U$ into the magic basis as $v$ and $u$ and choose $\alpha$, $\beta$, $\delta$ such that the eigenvalues of $v v^\tr$ match those of $u u^\tr$. (We include a global phase $e^{-i \pi/4}$ in $V$ to make it an element of SU(4).) This is done by comparing the analytical form of the eigenvalues of $v v^\tr$ to those of $u u^\tr$. Since $u u^\tr$ is unitary, it has complex eigenvalues of modulus one: $\lambda_j = e^{i \phi_j}$ ($j \in \{1, 2, 3, 4\}$). We find that $\alpha$, $\beta$, and $\delta$ are given by the means of pairs of eigenvalue phases. One possibility is $\alpha = (\phi_1+\phi_2)/2$, $\beta = (\phi_1+\phi_3)/2$, and $\delta = (\phi_2+\phi_3)/2$. Since no ordering of the eigenvalues is required, there are many such combinations that produce members of $U$'s local equivalence class. The proof of this assignment is by explicit calculation of the eigenvalues of $v v^\tr$ and is analogous to that given in ref.~\cite{shendePRA2004} for the controlled-NOT (CNOT) gate rather than our phase gate. 

Second, we find the four single-qubit rotations $A, B, C, D$ that comprise the remaining 12 degrees of freedom. Note that $v v^\tr$ and $u u^\tr$ are unitary symmetric matrices and therefore have real, orthonormal eigenvectors\cite{krausPRA2001,makhlinQIP2002}. Because they share eigenvalues, it is possible to simultaneously diagonalize them with matrices $k$ and $l$ such that
\begin{equation}
	kvv^\tr k^\tr = luu^\tr l^\tr .
	\label{eq:kvvkluul}
\end{equation}
Here, $k$ and $l$ are eigenvector matrices whose columns have been permuted such that equation~(\ref{eq:kvvkluul}) is valid. They are both members of SO(4) (if necessary, one of the eigenvectors can be negated to change the matrix determinant from -1 to 1). By rearranging equation~(\ref{eq:kvvkluul}), we obtain
\begin{equation}
	I = v^\dagger k^\tr l u u^\tr l^\tr k v^* = v^\dagger k^\tr l u \left(v^\dagger k^\tr l u\right)^\tr
\end{equation}
($I$ is the identity matrix), from which we define $m \equiv v^\dagger k^\tr l u$, also in SO(4). We thus have that
\begin{equation}
	u = l^\tr k v m
	\label{eq:magiccircuit}
\end{equation}
where $l^\tr k$ and $m$ are both real and in SO(4). Since they are real orthogonal matrices in the magic basis, they represent single-qubit rotations. We transform equation~(\ref{eq:magiccircuit}) into the computational basis and compare it with equation~(\ref{eq:computationalcircuit}) to find
\begin{eqnarray}
	C \otimes D = \Lambda\left(l^\tr k\right)\Lambda^\dagger \\
	A \otimes B = \Lambda m \Lambda^\dagger .
\end{eqnarray}
To finish, we split $A \otimes B$ and $C \otimes D$ into $A, B, C, D \in \textrm{SU(2)}$ and solve for $\theta$, $\phi$, and $\phi_z$ for each.



\bibliography{../../../../papers/hanneke}



\begin{list}{}{%
        \setlength{\leftmargin}{0in}%
        \setlength{\rightmargin}{0in}
        \setlength{\parsep}{0in}%
        \setlength{\itemsep}{0in}%
    }
 \item \textbf{Acknowledgements} This work was supported by DARPA, NSA, IARPA, and the NIST Quantum Information Program. We thank E. Knill for discussions and M. Biercuk and B. Eastin for comments on the manuscript. This paper is a contribution by the National Institute of Standards and Technology and not subject to US copyright.
 \item \textbf{Author Information} Correspondence and requests for materials should be addressed to D.H.~(email: SU4@david.hanneke.us).
\end{list}

\onecolumngrid
\setcounter{figure}{0}
\renewcommand{\thetable}{\arabic{table}}
\renewcommand{\tablename}{Supplementary Table}	
\renewcommand{\figurename}{Supplementary Figure}
\section*{Supplementary Information}

\begin{figure}[!b]
	\begin{center}
	\includegraphics[width=\textwidth]{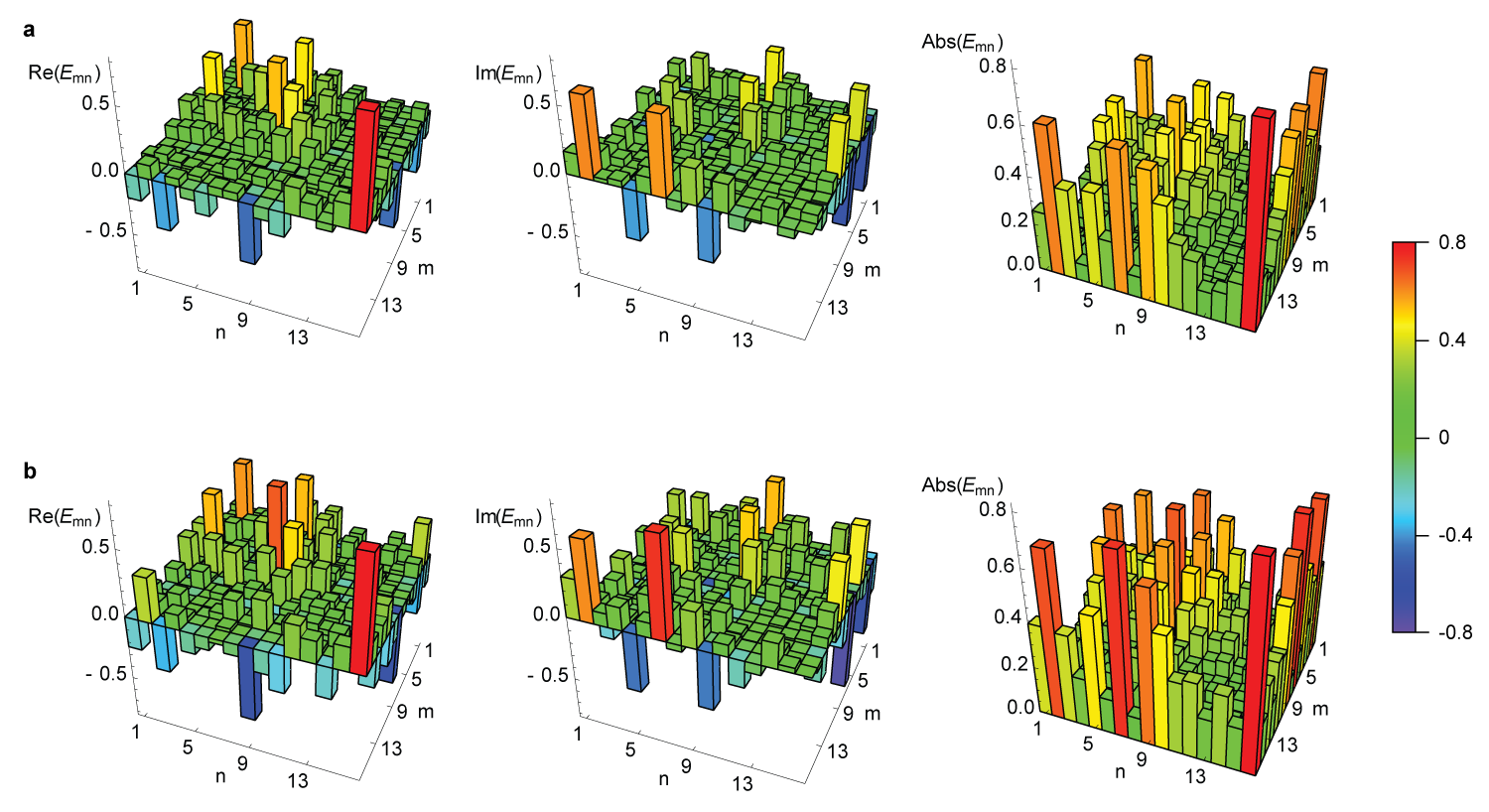}
	\end{center}
	\caption{Comparison of \textbf{a,} experimental and \textbf{b,} ideal process matrices of $U$; the fidelities are $F=0.77$ and $\bar{f}=0.82$. The process matrix $E$ contains the maps $\mathcal{E}(\ket{i}\bra{j})$ for all 16 elements $\ket{i}\bra{j}$. Since each map yields a $4\times4$ matrix $\mathcal{E}_{k,l}$, $E$ is a $16\times16$ matrix labeled by $m=4(i-1)+k$, $n=4(j-1)+l$. For example, the $\ket{\lowstate\lowstate}\bra{\highstate\highstate}$ $(i=4, j=1)$ element of an input density matrix is mapped to $\mathcal{E}(\ket{\lowstate\lowstate}\bra{\highstate\highstate})$, a $4\times4$ block of $E$ given by $m\in[13,16]$ and $n\in[1,4]$.}
\end{figure}

\subsection{\textsf{Example decomposition}}
As explicit examples of the decomposition described in the main text and Methods, Supplementary Table~1 gives two of the many possible decompositions of the unitary matrix
\begin{equation*} 
	U = \begin{pmatrix} -0.342+0.260i & -0.042-0.531i & -0.690+0.125i &  0.196-0.034i \\
											 0.195+0.743i & -0.138+0.090i & -0.064-0.520i & -0.304+0.127i \\
											-0.416+0.007i & -0.244+0.786i & -0.332+0.012i &  0.179+0.090i \\
											 0.074+0.217i & -0.084-0.076i &  0.320-0.145i &  0.871+0.229i \end{pmatrix}.
\end{equation*}
The process matrix from our implementation of the second of these is shown in Supplementary Figure~1.
\vspace{16pt}

\begin{table}[h]
\sffamily\renewcommand{\baselinestretch}{1.3}\scriptsize
\begin{center}
%

\begin{tabular}{cccccccccccccccc}
	\hline\hline
	$\alpha$ & $\beta$ & $\delta$ & $\theta_\textsf{A}$ & $\phi_\textsf{A}$ & $\phi_\textsf{z,A}$ & $\theta_\textsf{B}$ & $\phi_\textsf{B}$ & $\phi_\textsf{z,B}$  &	$\theta_\textsf{C}$ & $\phi_\textsf{C}$ & $\phi_\textsf{z,C}$ & $\theta_\textsf{D}$ & $\phi_\textsf{D}$ & $\phi_\textsf{z,D}$ & global phase \\
	\hline
	5.058 	 & 1.477 	 & 6.144 		& 4.165 		 & 4.759 		& 1.151 			 & 4.327 			& 5.678 	 & 2.088 &	0.856 		 & 5.210 		& 3.046 			 & 2.526 			& 4.528 	 & 1.570 				& $-1$ \\
	1.917 	 & 3.280 	 & 1.665 		& 1.254 		 & 2.987 		& 2.716 			 & 5.098 			& 2.537 	 & 0.693 &	5.438 		 & 1.068 		& 0.213 			 & 5.327 			& 0.062 	 & 2.838 				& $-i$ \\
	\hline\hline
\end{tabular}

	\caption{Two decompositions of the matrix $U$ shown above. The parameters refer to those shown in Figure~1b of the main text. The last column indicates an overall multiplicative phase factor to make the composite transformation mathematically equal to $U$, but it has no physical meaning.}
\end{center}
\end{table}

\subsection{\textsf{Details for figure 2}}

Figure 2 in the main text shows four examples of the output states formed while characterising the quantum processor. Supplementary Tables~2--3 give the input product states and 15 control parameters used to generate them. The unitary transformations decomposed in Supplementary Table~3 and used to create the states in Figure~2 are as follows.

\begin{equation*} 
	U_a = \begin{pmatrix} -0.362-0.428i & -0.501-0.125i &  0.054+0.083i & -0.464-0.440i \\
											   0.396+0.534i & -0.651-0.325i & -0.112-0.083i & -0.088+0.051i \\
											  -0.255+0.350i & -0.224+0.310i &  0.320+0.328i &  0.529-0.421i \\
											  -0.002+0.238i &  0.204+0.130i & -0.744+0.456i & -0.208-0.285i \end{pmatrix}
\end{equation*}
\begin{equation*} 
	U_b = \begin{pmatrix}  0.138+0.564i & -0.385-0.300i & -0.055+0.217i & -0.611+0.004i \\
												-0.196+0.035i & -0.256-0.388i & -0.576-0.027i &  0.386+0.512i \\
												-0.054-0.147i & -0.146-0.622i &  0.706+0.113i &  0.226+0.075i \\
												-0.152+0.758i &  0.324+0.180i &  0.277-0.171i &  0.289+0.273i \end{pmatrix}
\end{equation*}
\begin{equation*} 
	U_c = \begin{pmatrix} -0.645+0.445i & -0.154+0.006i & -0.174+0.561i &  0.127+0.037i \\
												-0.528+0.217i & -0.145-0.151i &  0.404-0.676i & -0.050-0.092i \\
												 0.012+0.136i &  0.534-0.495i & -0.183-0.016i &  0.250-0.596i \\
												-0.079+0.187i &  0.602-0.201i &  0.037+0.031i & -0.545+0.507i \end{pmatrix}
\end{equation*}
\begin{equation*} 
	U_d = \begin{pmatrix} -0.162+0.425i & -0.043+0.148i &  0.427+0.024i &  0.031-0.765i \\
												 0.091-0.304i & -0.614-0.158i & -0.372-0.433i &  0.104-0.401i \\
												 0.256-0.229i &  0.105+0.743i &  0.209-0.369i &  0.365+0.076i \\
												-0.607+0.454i & -0.079+0.072i & -0.116-0.546i &  0.035+0.319i \end{pmatrix}
\end{equation*}

\begin{table}[h]
\sffamily\renewcommand{\baselinestretch}{1.3}\scriptsize
\begin{center}
\begin{tabular}{ccc}
	\hline\hline
	subfigure & qubit 1 & qubit 2 \\
	\hline
	\textbf{a} & $\left(\up - i \down\right)/\sqrt{2}$ & $\down$ \\
	\textbf{b} & $\left(\up - i \down\right)/\sqrt{2}$ & $\down$ \\
	\textbf{c} & $\up$ & $\up$ \\
	\textbf{d} & $\left(\up - i \down\right)/\sqrt{2}$ & $\left(\up + \down\right)/\sqrt{2}$ \\
	\hline\hline
\end{tabular}

	\caption{Input states used to produce the states in Figure~2 of the main text.}
\end{center}
\end{table}

\begin{table}[h]
\sffamily\renewcommand{\baselinestretch}{1.3}\scriptsize
\begin{center}
\begin{tabular}{cccccccccccccccc>{\centering\hspace{0pt}}m{1 cm}}
	\hline\hline
	subfigure & $\alpha$ & $\beta$ & $\delta$ & $\theta_\textsf{A}$ & $\phi_\textsf{A}$ & $\phi_\textsf{z,A}$ & $\theta_\textsf{B}$ & $\phi_\textsf{B}$ & $\phi_\textsf{z,B}$  &	$\theta_\textsf{C}$ & $\phi_\textsf{C}$ & $\phi_\textsf{z,C}$ & $\theta_\textsf{D}$ & $\phi_\textsf{D}$ & $\phi_\textsf{z,D}$ & global phase \tabularnewline
	\hline
	\textbf{a} & 2.286 & 0.840 & 6.094 & 1.648 & 1.402 & 2.443 & 1.277 & 2.627 & 2.930 & 5.305 & 2.593 & 3.873 & 4.501 & 2.938 & 3.643 & $i$ \tabularnewline
	\textbf{b} & 6.196 & 1.439 & 1.119 & 2.542 & 3.079 & 4.151 & 1.982 & 1.744 & 6.140 & 0.833 & 2.349 & 1.947 & 3.742 & 0.011 & 1.487 & $-i$ \tabularnewline\
	\textbf{c} & 1.589 & 5.129 & 1.721 & 1.721 & 1.714 & 3.344 & 4.911 & 0.249 & 1.107 & 0.470 & 0.918 & 0.339 & 1.632 & 0.545 & 1.607 & $i$ \tabularnewline
	\textbf{d} & 0.389 & 0.750 & 1.392 & 5.423 & 1.300 & 6.004 & 4.824 & 1.961 & 3.199 & 4.351 & 1.100 & 1.501 & 2.352 & 2.022 & 1.426 & $-i$ \tabularnewline
	\hline\hline
\end{tabular}

	\caption{Single-qubit-gate parameters used for programming the quantum processor to produce the states in Figure~2 of the main text.}
\end{center}
\end{table}

\end{document}